\documentclass[useAMS,usenatbib,usegraphicx]{mn2e}
\usepackage{aas_macros}

\begin{document}

\title[memory and pulsar timing arrays]
{Gravitational-wave memory and pulsar timing arrays.}

\author[van Haasteren and Levin ]{Rutger van Haasteren$^1$ and Yuri Levin$^{1,2}$
\\
$^1$Leiden Observatory, P.O. Box 9513, NL-2300 RA Leiden
\\
$^2$Lorentz Institute, P.O. Box 9506, NL-2300 RA Leiden}

\date{printed \today}

\maketitle

\begin{abstract}
  Pulsar timing arrays (PTAs) are designed to detect gravitational waves with
  periods from several months to several years, e.g.~those produced by by wide
  supermassive black-hole binaries in the centers of distant galaxies.  Here we
  show that PTAs are also sensitive to mergers of supermassive black holes.
  While these mergers occur on a timescale too short to be resolvable by a PTA,
  they generate a  change of metric due to non-linear gravitational-wave memory
  which persists for the duration of the experiment and could be detected. We
  develop the theory of the single-source detection by PTAs, and derive the
  sensitivity of PTAs to the gravitational-wave memory jumps. We show that
  mergers of $10^8M_{\odot}$ black holes are $2-\sigma$-detectable (in a direction,
  polarization, and time-dependent way) out to co-moving distances of $\sim 1$
  billion light years.  Modern prediction for black-hole merger rates
  imply marginal to modest chance of an individual jump detection by
  currently developed PTAs. The sensitivity is expected to be somewhat higher for futuristic
  PTA experiments with SKA.
\end{abstract}

\section{introduction}
  Bursts of gravitational waves leave a permanent imprint on spacetime by
  causing a small permanent change of the metric, as computed in the transverse
  traceless gauge \citep{Payne1983, Christodoulou1991, Blanchet1992,
  Thorne1992}.  This gravitational-wave  ``memory jumps'' are particularly
  significant in the case of merger of a binary black hole, as was recently
  pointed out by \citet[hereafter F09]{Favata2009}.  Favata has shown (see
  Figure 1 of F09) that for the case of an equal-mass binary, a metric memory
  jump $\delta h$ was of the order of $\sim 5$ percent of $M/R$, where $M$ is
  the mass of the binary component and $R$ is the co-moving distance to the
  binary measured at redshift $0$ (hereafter $M$ is expressed in the geometric
  units, i.e.~$M=GM/c^2$). Furthermore, Favata has argued that the memory jumps
  were potentially detectable by LISA with high signal-to-noise ratio.  Favata's
  memory calculations make use of an approximate analytical treatment of the
  mergers, and need to be followed up with more definitive numerical
  calculations. Nevertheless, a number of analytical models explored in F09 show
  that the effect is clearly of high importance, and thus further investigations
  of detectability of the memory jumps are warranted. 

  Recently, there has been a renewed effort to measure gravitational waves from
  widely separated supermassive black-hole (SMBH) binaries by using precise
  timing of galactic millisecond pulsars \citep{Jenet2005, Manchester2006}.  In
  this paper we investigate whether pulsar timing arrays (PTAs) could be
  sensitive to the  memory jumps from physical mergers of the SMBHs at the end
  of the binary's life.  We demonstrate that modern PTAs \citep{Manchester2006},
  after $~10$ years of operation, will be sensitive to mergers of
  $10^8M_{\odot}$ black holes out to $\sim$billion light years; however the 
  chances of actual detection are small.  Futuristic PTA
  experiments, like those performed on the Square Kilometer Array
  \citep{Cordes2005}, offer a somewhat better prospect for the direct detection of
  gravitational-wave memory jumps.

\section{the signal}
  The gravitational waveform from a merger of SMBH pair consists of an {\it
  ac}-part and a {\it dc}-part; see Figure 1 of F09. The {\it ac}-part is
  short-period and short-lived, and hence is undetectable by a PTA. The {\it
  dc}-part is the gravitational-wave memory; it grows rapidly during the merger,
  on the timescale of $\sim 10M (1+z)\simeq 10^4 (M/10^8 M_{\odot})(1+z)\hbox{s}$,
  where $M$ is the mass of the SMBHs (assumed equal) and $z$ is the redshift of
  the merger.  After the burst passes, the change in metric persists, and as we
  explain below, it is this durable change in the metric that makes the main
  impact on the timing residuals.  Realistic PTA programs are designed to clock
  each of the pulsars with $\sim 2$-week intervals \citep[Bailes, private
  communications]{Manchester2006}.  Therefore, for $M=10^8 M_{\odot}$ SMBHs the
  growth of the memory-related metric change is not time-resolved by the timing
  measurements. Moreover, even for $M=10^{10} M_{\odot}$ SMBHs this growth
  occurs on the timescale much shorter than the duration on the experiment. We
  are therefore warranted to treat the {\it dc}-part of the gravitational wave
  as a discontinuous jump propagating through space,
  \begin{equation}
    h(\vec{r}, t)=h_0\times \Theta\left[(t-t_0)-\vec{n}\cdot \vec{r}\right],
    \label{dcwave}
  \end{equation}
  where $h_0$ is the amplitude of the jump, of the order of $0.05 M/R$,
  $\Theta(t)$ is the Heavyside function, $t_0$ is the moment of time when the
  gravitational-wave burst passes an observer, $\vec{r}$ is the location in
  space relative to the observer, and $\vec{n}$ is the unit vector pointed in
  the direction of the wave propagation. Here and below we set $c=1$.  We have
  used the plane-wave approximation, which is justified for treating
  extragalactic gravitational waves as they propagate through the Galaxy.

  For a single pulsar, the frequency of the pulse-arrival $\nu$ responds to a
  plane gravitational wave according to the following equation
  \citep{Estabrook1975, Hellings1983}:
  \begin{eqnarray}
    {\delta \nu(t)\over \nu}&=&B(\theta, \phi)\times
    \left[h(t)-h(t-r-r\cos\theta)\right],
    \label{deltanu}
  \end{eqnarray}
  where 
  \begin{equation}
    B(\theta,\phi)={1\over 2} \cos (2\phi) \left( 1-\cos\theta\right).
    \label{B}
  \end{equation}
  Here $r$ is the Earth-pulsar distance at an angle $\theta$ to the direction of
  the wave propagation, $\phi$ is the angle between the wave's principle
  polarization and the projection of the pulsar onto the plane perpendicular to
  the propagation direction, and $h(t)$ is the gravitational-wave strain at the
  observer's location.  Substituting Eq.~(\ref{dcwave}) into the above equation,
  we obtain the mathematical form of the signal:
  \begin{equation}
    {\delta \nu(t)\over \nu}=h_0 B(\theta, \phi)\times \left[ \Theta(t-t_0)-\Theta(t-t_1)\right],
    \label{deltanu1}
  \end{equation}
  where $t_1=t_0+r(1+\cos\theta)$. Thus the memory jump would cause a pair of
  pulse frequency jumps of equal magnitude and the opposite sign, separated by
  the time interval $r(1+\cos\theta)$. Since typical PTA pulsars are at least
  $\sim 10^3$ light years away, a single merger could generate at most one of
  the frequency jumps as seen during the $\sim 10$ years of a PTA experiment.
  The timing residuals from a single jump at $t=t_0$ are given by
  \begin{equation}
     m(t)=B(\theta,\phi)h_0\times \Theta(t-t_0)\times (t-t_0).
    \label{jump}
  \end{equation}
  For a single pulsar the frequency jump is indistinguishable from a fast
  glitch, and therefore single-pulsar data can only be used for placing upper
  limits on gravitational-wave memory jumps. The situation would be different
  for an array of pulsars, where simultaneous pulse frequency jumps would occur
  in  all of them at the time $t=t_0$ when the gravitational-wave burst would
  reach the Earth. Therefore a PTA could in principle be used to to detect
  memory jumps.

\section{Single-source detection by PTAs.}
  In this section we develop a mathematical framework for the single-source
  detection by a PTA. Our formalism is essentially Bayesian and follows closely
  the spirit of \citet[hereafter vHLML]{vanhaasteren2009}, although we will make
  a connection with the frequentist Wiener-filter estimator. We will then apply
  our general formalism to the memory jumps. The reader uninterested in
  mathematical details should skip the following subsection and go straight to
  the results in section~\ref{sec:tests}.

  There is a large body of literature on the single-source detection in the
  gravitational-wave community \citep{Finn1992, Owen1996, Brady1998}.
  The techniques which have been developed so far are designed specifically for
  the interferometric gravitational-wave detectors like LIGO and LISA.  There
  are several important modifications which need to be considered when applying
  these techniques to PTAs, among them\newline
  1. {\bf Discreteness of the data set.} A single timing residual per observed
  pulsar is obtained during the observing run; these runs are separated by at
  least several weeks. This is in contrast to the continuous (for all practical
  purposes) data stream in LIGO and LISA.\newline
  2. {\bf Subtraction of the systematic corrections.} The most essential of
  these is the quadratic component of the timing residuals due to pulsar
  spindown, but there may be others, e.g.~jumps of the zero point due to
  equipment change, annual modulations, etc.\newline
  3. {\bf Duration of the signal may be comparable to the duration of the
  experiment.} This is the case for both cosmological stochastic background
  considered in vHLML, and for the memory jumps considered here.  Thus frequency
  domain methods are not optimal, and time-domain formalism should be developed
  instead.   

  The Bayesian time-domain approach developed in vHLML and in this subsection is
  designed to tackle these 3 complications.
   
  Consider a collection of $N$ timing residuals $\delta t_{p}$ obtained from
  clocking a number of pulsars. Here $p$ is the composite index meant to
  indicate both the pulsar and the observing run together. Mathematically, we
  represent the residuals as follows:
  \begin{equation}
    \delta t_{p}=A\times s(t_{p})+\delta t^{n}_{p}+Q(t_{p}).
    \label{deltati}
  \end{equation}
  Here $s(t_{p})$ and $A$ are the known functional form and unknown amplitude of
  a gravitational-wave signal from a single source, $\delta t^{n}_p$ is the
  stochastic contribution from a combination of the timing and receiver noises,
  and 
  \begin{equation}
    Q(t_{p})=\Sigma_m\xi_mf_m(t_{p})
  \end{equation} 
  is the contribution from systematic errors of known functional forms
  $f_m(t_{p})$ but a-priory unknown magnitudes $\xi_m$. Below we shall specify
  $Q(t_{p})$ to be the unsubtracted part of the quadratic spindown, however for
  now we prefer to keep the discussion as general as possible.  We follow van
  Haasteren et al.~(2009) and rewrite Eq.~(\ref{deltati}) in a vector form:
  \begin{equation}
    \vec{\delta t}=A\vec{s}+\vec{\delta t}^n+F\vec{\xi}.
    \label{vector}
  \end{equation}
  Here the components of the column vectors $\vec{\delta t}$, $\vec{\delta
  t}^n$, $\vec{s}$, and $\vec{\xi}$ are given by $\delta t_p$, $\delta t^n_p$,
  $s(t_p)$, and $\xi_m$, and $F$ is a non-square matrix with the elements
  $F_{pm}=f_m(t_p)$. Henceforth we assume that $\delta t^{n}_{p}$ is the random
  Gaussian process, with the symmetric positive-definite coherence matrix $C$:
  \begin{equation}
    C_{pq}=<\delta t^n_{p} \delta t^n_{q}>.
    \label{cij}
  \end{equation}
  We can now write down the joint probability distribution for $A$ and $\xi_m$:
  \begin{eqnarray}
    P(A,\xi_m|\vec{\delta t})&=&{(1/M)}P_0(A,\xi_m)\times\label{jprob}\\
			 & & \exp\left[-{1\over 2}(\vec{\delta t}-A\vec{s}-F\vec{\xi})^{T}\times C^{-1}\times,\right.\nonumber\\
			 & &\left. (\vec{\delta t}-A\vec{s}-F\vec{\xi})\right].\nonumber
  \end{eqnarray}
  Here  $P_0(A,\xi_m)$ is the prior probability distribution, and $M$ is the
  overall normalization factor.  We now assume a flat prior $P_0(A, L_m)=const$,
  and marginalize over $\vec{\xi}$ in precisely the same way as shown in the
  Appendix of vHLML. As a result, we get the following Gaussian probability
  distribution for $A$:
  \begin{equation}
    P(A|\vec{\delta t})={1\over \sqrt{2\pi}\sigma}\exp\left[-{(A-\bar{A})^2\over 2\sigma^2}\right].
    \label{pA}
  \end{equation}
  Here, the mean value $\bar{A}$ and the standard deviation $\sigma$ are given by
  \begin{equation}
    \bar{A}={\vec{s}^{T}C^{\prime}\vec{\delta t}\over \vec{s}^{T}C^{\prime}\vec{s}},
    \label{barA}
  \end{equation}
  and 
  \begin{equation}
    \sigma=\left(\vec{s}^{T}C^{\prime}\vec{s}\right)^{-1/2},
    \label{sigma}
  \end{equation}
  where
  \begin{equation}
    C^{\prime}=C^{-1}-C^{-1}F\left(F^{T}C^{-1}F\right)^{-1}F^{T}C^{-1}.
    \label{Cprime}
  \end{equation}
  It is instructive and useful to re-write the above equations by introducing an
  inner product $\langle \vec{x},\vec{y}\rangle$ defined as 
  \begin{equation}
    \langle \vec{x}, \vec{y}\rangle=\vec{x}^{T}C^{-1}\vec{y}.
    \label{innerproduct}
  \end{equation}
  Let us choose an orthonormal basis\footnote{This is always possible by
  e.g.~the Gramm-Schmidt procedure.} $\hat{f}_i$ in the subspace spanned by
  $\vec{f}_m$, so that $\langle \hat{f}_i, \hat{f}_j \rangle=\delta_{ij}$.  We
  also introduce a projection operator
  \begin{equation}
    R=1-\Sigma_m \left|\hat{f}_m\rangle\langle \hat{f}_m\right|,
    \label{Proj}
  \end{equation}
  so that $R\vec{x}=\vec{x}-\Sigma_m\langle \hat{f}_m, \vec{x}\rangle \hat{f}_m$.
  All the usual identities for projection operators are satisfied, i.e.~$R^2=R$
  and $\langle R\vec{x}, R\vec{y}\rangle=\langle\vec{x}, R\vec{y}\rangle$.  We
  can then write
  \begin{equation}
    \bar{A}={\langle \vec{s}, R\vec{\delta t}\rangle\over \langle \vec{s}, R\vec{s}\rangle},
    \label{barA1}
  \end{equation}
  and 
  \begin{equation}
    \sigma=\langle\vec{s}, R\vec{s}\rangle^{-1/2}.
    \label{sigma1}
  \end{equation}
  If there are no systematic errors that need to be removed, than $R=1$ and the
  Eqs (\ref{barA1}) and (\ref{sigma1}) represent the time-domain version of the
  Wiener-filter estimator.

  \subsection{Other parameters}
    So far we have assumed that the gravitational-wave signal has a known
    functional form but unknown amplitude, and have explained how to measure or
    constrain this amplitude.  In reality, the waveform $\vec{s}(\vec{\eta},
    t_P)$ will depend on a number of a-priori unknown parameters $\vec{\eta}$,
    such as the starting time of the gravitational-wave burst and the direction
    from which this burst has come. These parameters enter into the probability
    distribution function through $\vec{s}$ in Eq.~(\ref{jprob}), and generally
    their distribution functions have to be estimated numerically. The estimates
    can be done via the matched filtering \citep{Owen1996} or by performing
    Markov-Chain Monte-Carlo (MCMC) simulations.  In section \ref{sec:tests}, we
    will demonstrate an example of an MCMC simulation for the memory jump. In
    this section, we show how to estimate an {\it average} statistical error on
    $\vec{\eta}$ for signals with high signal-to-noise ratios.

    Let us begin with a joint likelihood function for the amplitude $A$ and
    other parameters $\eta$:
    \begin{equation}
      L(A,\vec{\eta})=-(1/2)\langle A\vec{s}(\vec{\eta})-\vec{\delta t},  R (A\vec{s}(\vec{\eta})-\vec{\delta t})\rangle +Const.
      \label{likelyhood1}
    \end{equation}
    We now fix $A$ to its maximum-likelihood value $\langle \vec{s}(\vec{\eta}),
    R \vec{\delta t}\rangle/\langle \vec{s}(\vec{\eta}), R
    \vec{s}(\vec{\eta})\rangle$, and average over a large number of statistical
    realizations of the noise $\vec{\delta t}^n$.  The so-averaged likelihood
    function is given by
    \begin{eqnarray}
      L_{\rm av}(\vec{\eta})&=&-{(1/2)A_{t}^2\over \langle \vec{s}(\vec{\eta}), R\vec{s}(\vec{\eta})\rangle}\left[\langle\vec{s}(\vec{\eta}_t), R
		   \vec{s}(\vec{\eta}_t)\rangle \langle\vec{s}(\vec{\eta}), R
		   \vec{s}(\vec{\eta})\rangle-\right.\nonumber\\
		    & &\left.\langle\vec{s}(\vec{\eta}_t), R
		   \vec{s}(\vec{\eta})\rangle^2\right],
      \label{Leta}
    \end{eqnarray}
    where $A_t, \vec{\eta}_t$ are the true values for the signal present in all
    data realizations. We have omitted the additive constant.

    The expression in the square bracket is positive-definite, and $L_{\rm av}$
    is quadratic in $\vec{\eta}-\vec{\eta}_t$ for the values of $\vec{\eta}$
    close to the true values,
    \begin{equation}
      L_{\rm av}(\vec{\eta})\simeq-(1/2)(\vec{\eta}-\vec{\eta}_t)G (\vec{\eta}-\vec{\eta}_t),
      \label{fisher}
    \end{equation}
    where $G$ is the positive-definite Fisher information matrix. Its elements
    can be expressed as
    \begin{eqnarray}
      G_{ij}&=&{A_t^2/\langle\vec{s}, R\vec{s}\rangle}\left[\langle\vec{s}, R\vec{s}\rangle\langle{\partial \vec{s}
	      \over \partial\eta_i}, R{\partial \vec{s}\over \partial \eta_j}\rangle-\right.\nonumber\\
	   & & \left.\langle \vec{s}, R{\partial \vec{s}
	      \over \partial\eta_i}\rangle \langle \vec{s}, R{\partial \vec{s}
	      \over \partial\eta_j}\rangle\right],
      \label{fisher1}
    \end{eqnarray}
    evaluated at $\eta=\eta_t$.  The inverse of $G$ specifies the average error
    with which parameters $\vec{\eta}$ can be estimated from the data.          

\section{detectability of memory jumps}
  We now make an analytical estimate for detectability  of the memory jumps.
  For simplicity, we assume that  all of the pulsar observations are performed
  regularly so that  the timing-residual measurements are separated by a fixed
  time $\Delta t$, and that the whole experiment lasts over the time interval
  $[-T, T]$ (expressed in this way for mathematical convenience). Furthermore,
  we  assume that the timing/receiver noise is white, i.e. that for a pulsar $a$ 
  \begin{equation}
    \langle \delta t^n_i  \delta t^n_j\rangle=\sigma_a^2\delta_{ij}.
    \label{whitenoise}
  \end{equation}
  This assumption is probably not valid for some of the millisecond pulsars
  (Verbiest et al., in prep., van Haasteren et al., in prep.). We postpone
  discussion of the non-white noises to future work.

  To keep our exposition transparent, we consider the case when the array
  consists of a single pulsar $a$; generalization to several pulsars is
  straightforward and is shown later this section. Finally, we assume that the
  systematic error $Q(t_i)$ comprises only an unsubtracted component of the
  quadratic spindown,
  \begin{equation}
    Q(t_i)=A_0+A_1 t_i+A_2 t_i^2,
    \label{quad}
  \end{equation}
  where $A_0$, $A_1$, and $A_2$ are a-priori unknown parameters.

  We now come back to the formalism developed in the previous section. The inner
  product defined in Eq.~(\ref{innerproduct}) takes a simple form:
  \begin{eqnarray}
    \langle \vec{x}, \vec{y}\rangle&=&{1\over \sigma_a^2}\sum_i x(t_i)y(t_i)\label{innerproduct1}\\
				    &\simeq& {1\over \sigma_a^2\Delta t}\int_{-T}^Tx(t)y(t)dt,\nonumber
  \end{eqnarray}
  where we have assumed $\Delta t\ll T$ and have substituted the sum with the
  integral in the last equation.  We now choose orthonormal basis vectors
  $\hat{f}_{1,2,3}(t)$ which span the linear space of quadratic functions:
  \begin{eqnarray}
    \hat{f}_1(t)&=&\sigma_a\sqrt{\Delta t\over T} {1\over \sqrt{2}}\label{basis2}\\
    \hat{f}_2(t)&=&\sigma_a\sqrt{\Delta t\over T} \sqrt{3\over 2} {t\over T}\nonumber\\
    \hat{f}_3(t)&=&\sigma_a\sqrt{\Delta t\over T}\sqrt{45\over 8}\left[\left({t\over T}\right)^2-
		   {1\over 3}\right].\nonumber
  \end{eqnarray}
  From Eq.~(\ref{jump}) the gravitational-wave induced timing residuals are given by
  $\delta t(t)=h_0 s(t)$, where
  \begin{equation}
     s(t)=B(\theta,\phi)\times \Theta(t-t_0)\times (t-t_0).
    \label{jump1}
  \end{equation}
   The expected measurement error of the jump amplitude $h_0$ is given by
   Eq.~(\ref{sigma1}):
  \begin{equation}
    \sigma_{h_0}=\left[\langle \vec{s},\vec{s}\rangle^2-\sum_{i=1,2,3}\langle \vec{s}, \hat{f}_i\rangle^2\right]^{-1/2}.
    \label{sigmah0}
  \end{equation}
  Substituting Eqs.~(\ref{innerproduct1}), (\ref{basis2}), and (\ref{jump1})
  into Eq.~(\ref{sigmah0}), one gets after some algebra:
  \begin{equation}
    \sigma_{h_0}={1\over B(\theta,\phi)} {\sigma_a\over T}\sqrt{48\over N p^3\left(1-{15\over 16}p\right)}.
    \label{sigmah00}
  \end{equation}
  Here $N=2T/\Delta t$ is the number of measurements, and
  \begin{equation}
    p=1-\left(t_0/T\right)^2.
    \label{p}
  \end{equation}
  For an array consisting of multiple pulsars, and with the assumption that the
  timing residuals are obtained for all of them during each of the $N$ observing
  runs, the above expression for $\sigma_{h_0}$ is modified as follows:
  \begin{equation}
    \sigma_{h_0}={\sigma_{\rm eff}\over T}\sqrt{48\over Np^3\left(1-{15\over 16}p\right)},
    \label{sigmah01}
  \end{equation} 
  where 
  \begin{equation}
    \sigma_{\rm eff}=\left[\sum_a\left(B^2(\theta_a,\phi_a)/\sigma_a^2\right)\right]^{-1/2}.
    \label{sigmaeff}
  \end{equation}
  Several remarks are in order:

  1. The error $\sigma_{h_0}$ diverges when $p=0$, i.e.~when $t_0=\pm T$. This
  is as expected: when the memory jump arrives at the beginning or at the end of
  the timing-array experiment, it gets entirely fitted out when the pulsar spin
  frequency is determined, and is thus undetectable. 

  2. Naively, one may expect the optimal sensitivity when the jump arrives
  exactly in the middle of the experiment's time interval, i.e.~when $t_0=0$.
  This is not so; the optimal sensitivity is achieved for $t_0/T=\pm 1/\sqrt{5}$
  when the error equals
  \begin{equation}
    \sigma_{h_0}={\sigma_{\rm eff}\over T}\sqrt{375\over N}.
    \label{sigmah02}
  \end{equation}

  3. The sky-average value for $B^2(\theta, \phi)$ is $1/6$. Therefore, for an
  array consisting of a large number of pulsars $N_p$ which are distributed in
  the sky isotropically and which have the same amplitude of timing/receiver
  noise $\sigma_a=\sigma$, the $\sigma_{\rm eff}$ in Eq.~(\ref{sigmah01})  is
  given by
  \begin{equation}
    \sigma_{\rm eff}=\sigma\sqrt{6/N_p}.
    \label{sigmaeff1}
  \end{equation}

  4. While the timing precision of future timing arrays is somewhat uncertain,
  it is instructive to consider a numerical example.  Lets assume $T=5$yr (i.e.,
  the 10-year duration of the experiment), $N=250$ (i.e., roughly bi-weekly
  timing-residual measurements), $N_p=20$ isotropically-distributed pulsars
  (this is the current number of clocked millisecond pulsars), and
  $\sigma_a=100$ns (this sensitivity is currently achieved for only several
  pulsars).  Then for optimal arrival time $t_0=\pm T/\sqrt{5}$, the array
  sensitivity is
  \begin{equation}
    \sigma_{h_0}=4.5\times 10^{-16}.
    \label{sigmah03}
  \end{equation}
  For a binary consisting of two black holes of the mass $M$, the memory jump is
  estimated in F09 to be
  \begin{equation}
    h_0=\eta {M\over R}\simeq 8\times 10^{-16}{\eta\over 0.05}\left({M\over 10^8M_{\odot}}\right)
					   \left({10^9\hbox{light-years}\over R}\right),
    \label{h01}
  \end{equation}
  where $\eta\sim 0.05$ is the direction-dependent numerical parameter. In this
  example, the pulsar-timing array is sensitive to the memory jumps from
  black-hole mergers at redshifts $z<z_0$, where $z_0\sim 0.1$ for
  $M=10^8M_{\odot}$, and $z_0\sim 1$ for $M=10^9M_{\odot}$.

  \subsection{Arrival time}
    It is possible to estimate the array's sensitivity to the memory-jump
    arrival time, $t_0$. We use Eqs.~(\ref{fisher1}) and (\ref{jump1}), and
    after some algebra\footnote{A useful identity: 
      \begin{equation}
	{\partial[(t-t_0)\Theta(t-t_0)]\over \partial t_0}=-\Theta(t-t_0).
      \end{equation}
    } 
    get
    \begin{equation}
      \sigma_{t_0}=T \left({h_0T\over \sigma_{\rm eff}}\right)^{-1}\sqrt{2/N}\chi(p),
      \label{sigmat0}
    \end{equation}
    where 
    \begin{equation}
      {1\over \chi^2(p)}={1\over 2}p\left(1+{5\over 4}p^2-2p\right)-{3(1-p)p\left[1-(5/8)p\right]^2
			      \over 2[1-(15/16)p]}.
    \end{equation}

  \subsection{Source position}
\begin{figure}
    \includegraphics[width=0.5\textwidth]{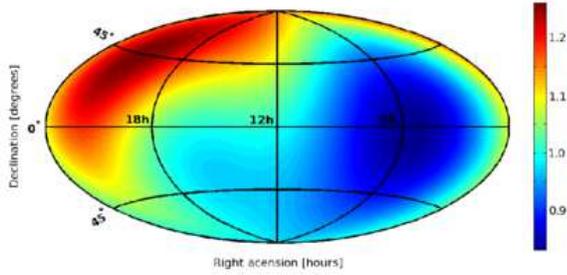}
    \caption{The relative sensitivity $\sigma_{\rm eff}$ for the pulsars of the
      European Pulsar Timing Array. The scaling has been chosen such that a
      value of $1$ indicates that the same sensitivity for that source position
      would have been achieved with a perfect isotropic PTA (i.e. $B^2 =
      \frac{1}{6}$)} \label{fig:eptasensitivity}
    \end{figure}

\begin{figure}
    \includegraphics[width=0.5\textwidth]{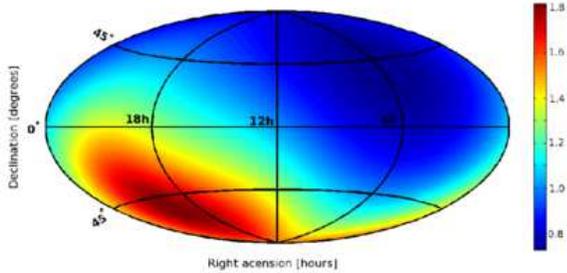}
    \caption{The relative sensitivity $\sigma_{\rm eff}$ for the pulsars of the
      Parkes Pulsar Timing Array. The scaling has been chosen such that a value
      of $1$ indicates that the same sensitivity for that source position would
      have been achieved with a perfect isotropic PTA (i.e. $B^2 =
      \frac{1}{6}$)} \label{fig:parkessensitivity}
    \end{figure}

    The array's sensitivity gravitational-wave memory is dependent on source
    position since the number and the position of the pulsars in current PTAs is
    not sufficient to justify the assumption of isotropy made in
    Eq.~(\ref{sigmaeff1}). We will therefore calculate the value of $\left[
    \sum_{a}B^2\left( \theta_{a},\phi_{a} \right) \right]^{-1/2}$ for current
    PTAs. Since the polarisation of the gravitational-wave memory signal is an
    unknown independent parameter, we average over the polarisation and obtain
    for the angular sensitivity:
    \begin{eqnarray}
      \sigma_{h_0}\left( \phi_{s},\theta_{s} \right) &\sim& \left[
      \sum_{a}B^2\left(\phi_{s},\theta_{s}, \theta_{a},\phi_{a} \right)
      \right]^{-1/2} \\
      &=& \left[ \sum_{a} \frac{1}{8}\left( 1+\cos\theta_{a}\left( \phi_{s},\theta_{s} \right)
      \right)^2 \right]^{-1/2}.
      \label{eq:angsens}
    \end{eqnarray}
    Here we have assumed that all pulsars have equal timing precision.
    $\phi_s$ and $\theta_s$ are the position angles of the gravitational-wave
    memory source, and $\theta_a$ is the polar angle of pulsar $a$ in a
    coordinate system with $\left( \phi_s,\theta_s \right)$ at the north-pole.
    In figure~\ref{fig:eptasensitivity} and~\ref{fig:parkessensitivity} the
    sensitivity to different gravitational-wave memory source positions is shown
    for respectively the European Pulsar Timing Array and the Parkes Pulsar
    Timing Array projects.

\section{Tests using mock data}
  \label{sec:tests}
  We test the array's sensitivity to gravitational-wave memory signals using
  mock timing residuals for a number of millisecond pulsars. In this whole
  section, all the mock timing residuals were generated in two steps:\newline
  1) A set of timing residuals was generated using the pulsar timing package
  tempo2 \citep{Hobbs2006}. We assume that the observations
  are taken
  tri-weekly over a time-span of $10$ years. The pulsar timing noise was set to
  $100$ ns white noise.\newline
  2) A gravitational-wave memory signal was added according
  to Eq.~(\ref{jump}), with a memory-jump arrival time set to be optimal for
  sensitivity: $\frac{t_{0}}{T} = \frac{1}{\sqrt{5}}$. The direction and
  polarisation of the gravitational-wave memory signal were chosen
  randomly - the coordinates happened to have declination $90^o$.\newline
  In the following subsections we describe tests which have fixed parameters for step 1,
  but systematically varied amplitude for step 2, and we use these tests
  to study the sensitivity of the array.

  \subsection{Used models}
    \label{sec:gwsensitivity}
    In principle, we would like to realistically extrapolate the results we obtain
    here for mock datasets to future real datasets from PTA projects. Several
    practical notes are in order to justify the models we use here to analyse
    the mock datasets:\newline
    1) From equation~(\ref{quad}) onward, we  assume that the 
   systematic-error contributions to the timing residuals  consist only of the
    quadratic spindown. In reality, pulsar observers must fit many model parameters 
    to the data, and have developed appropriate fitting routines within timing packages like tempo2.
    Similar to the quadratic spindown discussed in this paper, all the
    parameters of the timing model are linear or linearised in tempo2, and
    therefore those parameters are of known functional form. Since the
    subtraction of quadratic spindown decreases the sensitivity of the PTA to
    gravitational-wave memory signals, we would expect the same thing to be true
    for the rest of the timing model.\newline
    2) The error-bars on the pulse arrival time obtained from correlating the
    measured pulse-profile with the template of the pulse-profile 
     are generally not completely
    trusted. Many pulsar astronomers  invoke an extra ``fudge'' factor that
    adjusts the error-bars on the timing-residuals to make sure that the errors
    one gets on the parameters of the timing-model are not underestimated.
    Usually the ``fudge'' factor, which is known as an  $efac$ value is set to the
    value which makes the
    reduced $\chi^2$ of the timing solution to be equal to $1$.

    In order to check the significance of both limitations 1 \& 2, we perform the following test.
    We take a realistic set of pulsars with realistic timing models: the pulsar
    positions and timing models of the PPTA pulsars. We then simulate white
    timing-residuals and a gravitational-wave memory signal with amplitude
    $h_0=10^{-15}$, and we produce the posterior distribution of Eq.~(\ref{pA}) in
    three different ways: \newline
    a) We marginalise over only the quadratic functions of
    Eq.~(\ref{basis2}), which should yield the result of
    Eq.~(\ref{sigmah01}).\newline
    b) We marginalise over the all timing model parameters included in the
    tempo2 analysis when producing the timing-residuals.\newline
    c) We marginalise over all the timing model parameters, and we also
    marginalise over the efac values using the numerical techniques of vHLML. By
    estimating the efac value simultaneously with the gravitational-wave memory
    signal, we are able to completely separate the two effects. Note that this
    procedure will not destroy information about the relative size of the
    error-bars for timing-residuals of the same pulsar.

\begin{figure}
    \includegraphics[width=0.5\textwidth]{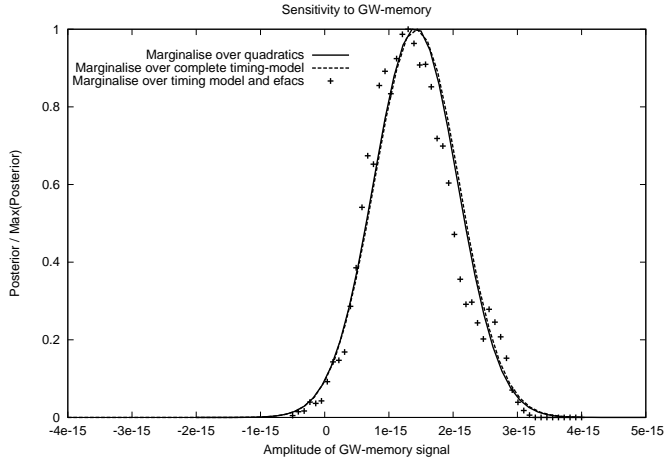}
    \caption{The posterior distribution of the gravitational-wave memory
      amplitude. The two solid lines are the result of an analysis where we only
      analytically marginalise over the full timing model or just quadratic
      spindown. The points are the result of a marginalisation over the full
      timing model and the efac values as well. From the Gaussians, the
      sensitivity can be reliably estimated at: $\sigma_{h_0} =
      \frac{FWHM}{2\sqrt{2\ln2}} = 6.5 \times 10^{-16}$}
      \label{fig:likelihood_amplitude}
    \end{figure}

    We present the result of this analysis in Figure
    \ref{fig:likelihood_amplitude}. Based on the $185$ observations per pulsar
    in the dataset and the direction of the gravitational-wave memory signal, we
    can calculate the theoretical sensitivity of the array using
    Eq.~(\ref{sigmah02}) and Eq.~(\ref{sigmaeff1}). This yields a value of:
    \begin{equation}
      \sigma_{h_0} = 6.4 \times 10^{-16}.
      \label{sigmah0parkestheory}
    \end{equation}
    We can also calculate this value for the three graphs in Figure
    \ref{fig:likelihood_amplitude}. The three graphs lie close enough on top of
    each other to conclude that one value applies to all three of them:
    \begin{equation}
      \sigma_{h_0} = 6.5 \times 10^{-16},
      \label{sigmah0parkesmock}
    \end{equation}
    which is in good agreement with the theoretical value. It appears that both
    note 1 and 2 mentioned above are not of great influence to the sensitivity
    of PTAs to gravitational-wave memory detection; the theoretical calculations
    of this paper are a good representation of the models mentioned in this
    section.

  \subsection{Upper-limits and detecting the signal}
    When there is no detectable gravitational-wave memory signal present
    in the data, we can set some upper-limit on the signal amplitude using
    the algorithm presented in this paper. Here we will analyse datasets with no
    or no fully detectable gravitational-wave memory signal in it, and a dataset
    with a well-detectable signal using the MCMC method of vHLML. We will
    calculate the marginalised posterior distributions for the $5$ parameters of
    the gravitational-wave memory signal. The interesting parameters in the case
    of an upper-limit are the amplitude and the arrival time of the jump. A
    marginalised posterior for those two parameters are then presented as
    two-dimensional posterior plots. Note that the difference with the analysis
    in section \ref{sec:gwsensitivity} is that we vary all gravitational-wave
    memory parameters, instead of only the amplitude. Note that we do
    marginalise over all the efac values as discussed in section
    \ref{sec:gwsensitivity}, unless stated otherwise.

\begin{figure}
    \includegraphics[width=0.5\textwidth]{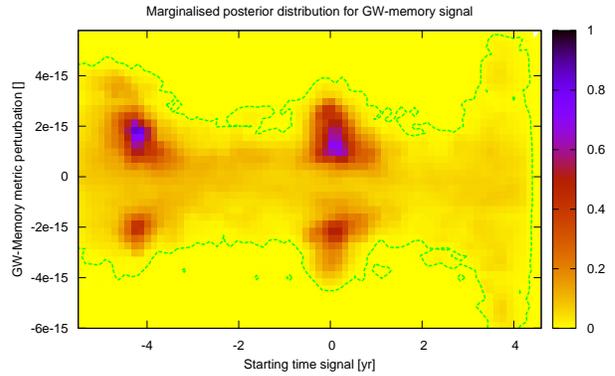}
    \caption{The marginalised posterior distribution for the gravitational-wave
      memory signal amplitude and arrival time of the jump. In this case a
      dataset was analysed that did not contain any gravitational-wave memory
      signal.}
      \label{fig:2dlikelihood_white}
    \end{figure}

    In Figure~\ref{fig:2dlikelihood_white} we show the result of an analysis of
    a dataset where we have not added any gravitational-wave memory signal to
    the timing-residuals. The $3-\sigma$ contour is drawn, which serves as an
    upper-limit to the memory amplitude. We see that we can exclude a
    gravitational-wave memory signal at $\frac{t_0}{T} = \frac{1}{\sqrt{5}}$ of
    amplitude $3 \times 10^{-15}$ and higher. We see that this value is over a
    factor of $4$ higher than what is predicted by
    Eq.~(\ref{sigmah0parkestheory}). This is to be expected, since:\newline
    1) We give a $3-\sigma$ limit here, instead of the $1-\sigma$
    sensitivity.\newline
    2) We also marginalise over the arrival time and other parameters of the
    memory signal, reducing the sensitivity.\newline
    Because of these reasons, we argue that the \emph{minimal} upper-limit one
    can set on the gravitational-wave memory signal using a specific PTA is the
    sensitivity calculated using Eq.~(\ref{sigmah0parkestheory}) multiplied by
    $4$.

\begin{figure}
    \includegraphics[width=0.5\textwidth]{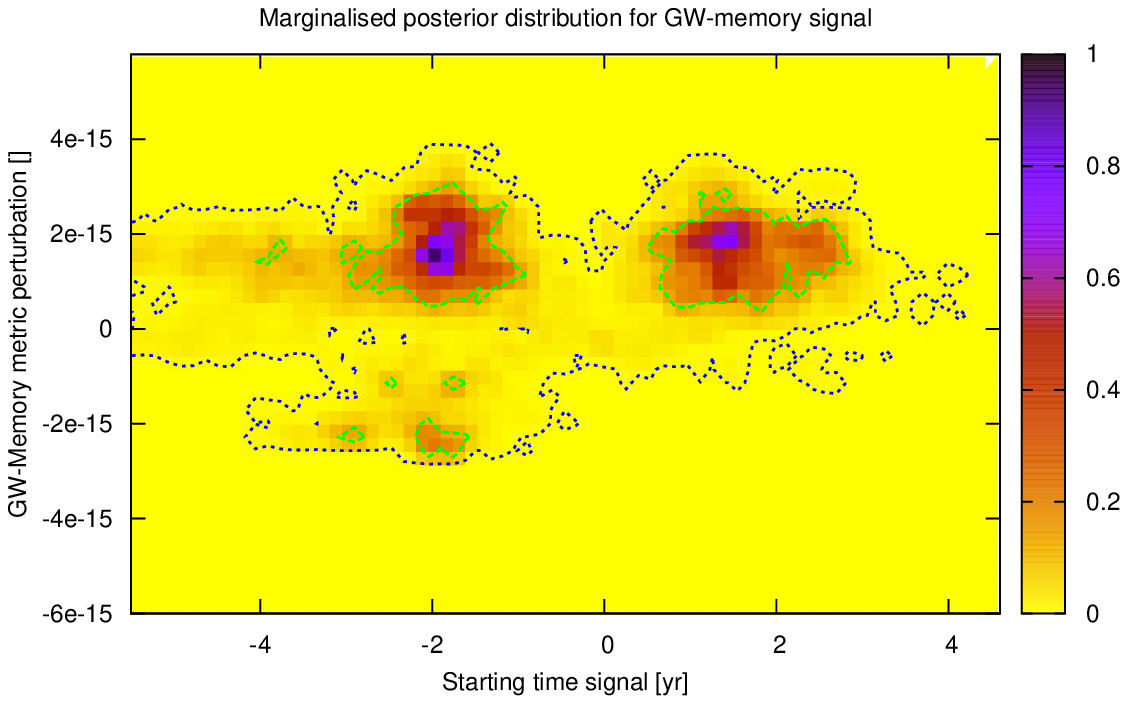}
    \caption{The marginalised posterior distribution for the gravitational-wave
      memory signal amplitude and arrival time of the jump. Here a
      gravitational-wave signal with an amplitude of $10^{-15}$ was added to the
      white residuals. The contour drawn is the $3-\sigma$ contour.}
      \label{fig:2dlikelihood_white_15}
    \end{figure}

\begin{figure}
    \includegraphics[width=0.5\textwidth]{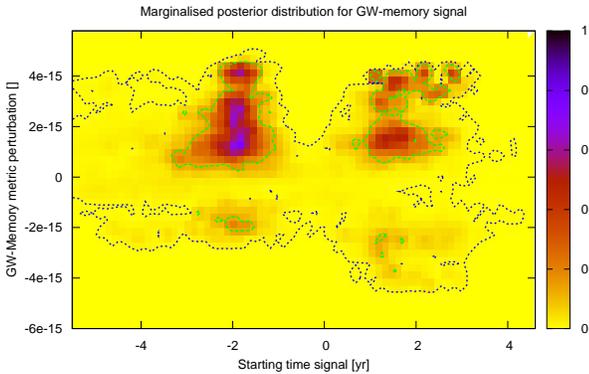}
    \caption{The marginalised posterior distribution for the gravitational-wave
      memory signal amplitude and arrival time of the jump. Here a
      gravitational-wave signal with an amplitude of $10^{-15}$ was added to the
      white residuals. This analysis has been done without marginalising over
      the efac values. The contour drawn is the $3-\sigma$ contour.}
      \label{fig:2dlikelihood_white_15_noefac}
    \end{figure}

    Next we produce a set of timing-residuals with a memory signal of amplitude
    $h_0=10^{-15}$. According the result mentioned above, the memory signal should
    not be resolvable with this timing precision. The result is shown in
    Figure~\ref{fig:2dlikelihood_white_15}. We see that we can indeed merely set
    an upper-limit again. In order to check the effect of marginalising over the
    efac values as mentioned in section~\ref{sec:gwsensitivity}, we also perform
    an analysis where we pretend we do know the efac values prior to the
    analysis. The result is shown in
    Figure~\ref{fig:2dlikelihood_white_15_noefac}. We see no significant
    difference between the two models.

\begin{figure}
    \includegraphics[width=0.5\textwidth]{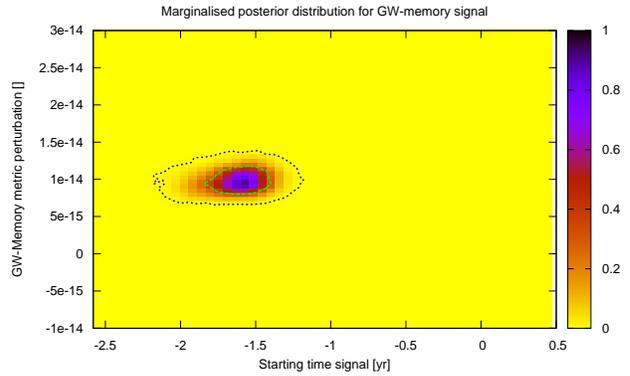}
    \caption{The marginalised posterior distribution for the gravitational-wave
      memory signal amplitude and arrival time of the jump. Here a
      gravitational-wave signal with an amplitude of $10^{-14}$ was added to the
      white residuals, indicated with a '+' in the figure. The contours drawn
      are the $1-\sigma$ and $3-\sigma$ contours.}
      \label{fig:2dlikelihood_white_14_ampstart}
    \end{figure}

\begin{figure}
    \includegraphics[width=0.5\textwidth]{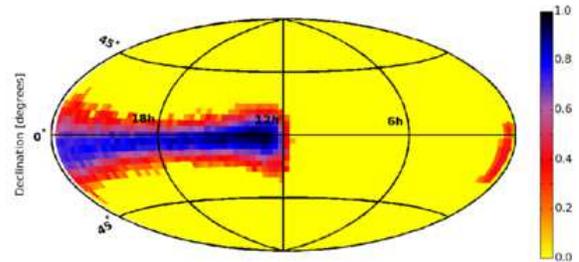}
    \caption{The marginalised posterior distribution for the sky location of the
      gravitational-wave memory signal. We can see here that we can marginally
      determine the direction of the source. The source positions used to
      generate the residuals were: (declination, right ascension) = ($90^0,
      12.4\hbox{hr}$).}
      \label{fig:2dlikelihood_white_14_pos}
    \end{figure}

    Finally, we also analyse a dataset with a gravitational-wave memory signal
    with an amplitude larger than the $3 \times 10^{-15}$ upper-limit of the
    white set mentioned above. Here we have added a memory signal with an
    amplitude of $10^{-14}$. In Figure~\ref{fig:2dlikelihood_white_14_ampstart}
    we see that we have a definite detection of the signal: if we consider the
    $3-\sigma$ contours, we see that we can restrict the gravitational-wave
    memory amplitude between $\left[ 6.6 \times 10^{-15}, 1.35 \times 10^{-14}
    \right]$. Again, this value is higher than the value predicted by
    Eq.~(\ref{sigmah0parkestheory}) due to us including more parameters in the
    model than just the memory amplitude. In
    Figure~\ref{fig:2dlikelihood_white_14_pos} we see that we can also reliably
    resolve the position of the source in this case.

\section{Discussion}

  In this paper, we have shown that  gravitational-wave memory signals from 
  SMBH binary mergers are in principle detectable by PTAs, and that 
  $2-\sigma$ constraints are possible on $M=10^8M_{\odot}$ mergers out
to redshift of $\sim 0.1$ (while those with $M=10^{10}M_{\odot}$ should 
  be detectable throughout the Universe). How frequently do these mergers
  occur during the PTA lifetime? Recent calculations of \citet[SVH]{Sesana2007}
 are not too encouraging. SVH compute, for several models of SMBH merger trees,
 the rate of SMBH mergers as
seen on Earth, as a function of mass (their figure 1d), as well as a multidute
of other parameters for these mergers. From their plots one infers few$\times 10^{-2}-10^{-3}$
PTA-observable mergers per year, which converts to at most $0.1-0.01$ detected
mergers during the PTA lifetime of $\sim 10$ years (NB: during the PTA existence,
only a fraction of time will spent near the arrival times with optimal sensitivity).
It is conceivable that SVH estimates are on the conservative side, since
the mergers of heavy black holes may be stalled (due to the ``last parsec'' problem)
and may occur at a significantly later time than the mergers of their host halos.
In this case, some fraction of high-redshift mergers may be pushed towards lower redshifts
and become PTA-detectable. Detailed calculations are needed to find out whether
this process could substantially increase the rate of PTA-detectable mergers. 
It is also worth pointing out that a futuristic PTA experiment based on a Square
Kilometer Array may attain up to an order of magnitude higher sensitivity that
the currently developed PTAs. 

The methods presented in this paper are useful beyond the particular application
that we discuss. The algorithm presented here is
  suitable for any single-source detection
  in general when  the gravitational waveform has known functional
form. Further applications will be presented elsewhere.

  \subsection{comparison with other work}
  When this paper was already finished, a preprint by \citet[PBP]{Pshirkov2009}
  has appeared on the arxiv which has carried out a similar analysis to the one
  presented here. Our expressions for the signal-to-noise ratio for the memory
  jump agree for the case of the white pulsar noise. PBPs treatment of cosmology
  is more detailed than ours, while the moderately pessimistic predicted
  detection rates are broadly consistent between the 2 papers.  Our method for
  signal extraction is more generally applicable than PBS's since it is
  optimized for any spectral type of pulsar noise, takes into consideration not
  just the signal magnitude but also other signal parameters, and is tested on
  mock data.

\section*{Acknowledgments}
  This research is supported by the Netherlands organisation for Scientific
  Research (NWO) through VIDI grant 639.042.607.


  \bibliographystyle{mn2e.bst}
  \bibliography{gwmemory-1}

\end{document}